\documentclass[prb,twocolumn,superscriptaddress,showpacs,preprintnumbers,amsmath,amssymb,floatfix]{revtex4-1}

\usepackage{graphicx,tabularx,multirow}
\usepackage{dcolumn}
\usepackage{bm}
\usepackage{upgreek,nicefrac}
\usepackage{color,soul}

\newcommand{\tio}{TiO$_2$}
\newcommand{\tioo}{TiO$_2$ }

\newcommand{\mye}[1]{Eq.~(\ref{eq:#1})}
\renewcommand{\hl}[1]{#1}

\begin{document}

\title{Electrical Transport Properties of Polycrystalline and
  Amorphous \tioo Single Nanotubes}

\author{Markus Stiller}
\email{markus@mstiller.org}

\author{Jose Barzola-Quiquia} 

\author{Pablo Esquinazi} 
\affiliation{Division of Superconductivity and
  Magnetism, Institute for Experimental Physics II, University of
  Leipzig, 04103 Leipzig, Germany}

\author{Seulgi So}
\author{Imgon Hwang}
\author{Patrik Schmuki}

\affiliation{Chair for Surface Science and Corrosion Department
  Material Science and Engineering, University of Erlangen, D-91058
  Erlangen, Germany}

\author{Julia B{\"o}ttner} \author{Irina Estrela-Lopis}
\affiliation{Institute of Medical Physics and Biophysics, University
  of Leipzig, 04107 Leipzig, Germany}

\date{\today}
\begin{abstract}
  The electrical transport properties of anodically grown \tioo
  nanotubes was investigated. Amorphous nanotubes were anodically
  grown on titanium foil and transformed through annealing into the
  anatase phase.  Amorphous and polycrystalline single nanotubes were
  isolated and contacted for measurements of the electrical
  resistance. Non-linear current-voltage characteristics were
  explained using the fluctuation induced tunneling conduction model.
  A clear enhancement of the conductance was induced in an insulating
  anatase nanotube through low-energy Ar/H ion irradiation. Confocal
  Raman spectroscopy shows that the annealed samples were in anatase
  phase and a blueshift due to phonon confinement was observed.
\end{abstract}
\maketitle

\section{Introduction}
\label{introduction}

Titanium dioxide nanotube arrays, formed by self-organizing
anodization, have attracted considerable attention.  These 1D
structures are used in wide range of applications such as electrodes
in catalysis~\cite{HK96}, photocatalysis~\cite{FH72}, dye-synthesized
solar cells~\cite{RG91}, gas sensors~\cite{SYKHSY09},
\hl{photoelectrochemical water splitting}~\cite{MG15b,MG15a}, in
batteries~\cite{ABSTS05} or for CO$_2$
reduction~\cite{WSRPRA10,VPLG09}. Biocompatible
Ti$_{46}$Si$_{12}$O$_{42}$ nanostructured surfaces can be used to
enhance cell attachment and proliferation~\cite{ZCYSWHGC16}. This
outstanding flexibility is a consequence of different electronic,
chemical and ionic properties of anatase, brookite and
rutile~\cite{D03,CM07}. \hl{Anatase is often more interesting for
  applications, such as solar cells, due to the larger electron
  mobility compared to rutile.}~\cite{TPSSL94} The combination with the
large active surface area of nanostructures, results in a variety of
dielectric, conducting, magnetic, catalytic and other physical and
chemical properties. \hl{Doping can be used to enhance the photo
  catalytic activity of} \tioo \hl{nanotubes, e.g.~with
  W}~\cite{MGG17,MG16a,MGD15}, \hl{Au/W}~\cite{MG16b},
\hl{Co}~\cite{MGG15,MA16}, \hl{or Co}~\cite{MG16a}.

The general mechanism of electron transport, particularly in
nanostructures of \tioo, is not well studied and understood. Most of
the published work focuses on macroscopic samples such as nanotubular
arrays~\cite{MVPSG06,LSCNZL05}, where parasite effects such as contact
contributions in two-point (2P) measurements or the scattering of
light inside the tangle of tubes, can have considerable influence on
the measurements. Depending on the contacts, non-metallic clusters
(Au), Schottky barriers (Pt) or oxide layers (Al, Cr) could be
formed~\cite{D03}. The magnetic~\cite{CC13,SQESMLG16} and electrical
transport properties of \tioo depend strongly on structural defects
(single crystal band gap: $3.0-3.2$~eV~\cite{STBGML94,PCM78}), such as
oxygen or titanium vacancies. The transport properties can also be
strongly influenced by the scattering on the sample surface, as was
found for polycrystalline macroscopic arrays of \tioo
nanotubes~\cite{M07}. In addition, it was shown that, intragrain and
intergrain conduction processes play an important role in disordered
nanowires and nanotubes~\cite{SQZE15,HandNanophysics}.  Therefore, a
large number of different resistivities have been reported,
e.g.~$10^4~\Omega$cm (2P, top/bottom tube contacts)~\cite{TRAHS10},
$10^{−2}\dots10^{−3}~\Omega$cm~\cite{SQLEKAS13} and
$1~\Omega$cm~\cite{FRPDAM10} using a four-point (4P) probes
method. For comparison, the obtained resistivity of different
polycrystalline bulk anatase covers a large range:
$10^2\dots10^7~\Omega$cm~\cite{TPSSL94,AKH06,HGZ04}.

Recently, fluctuation induced tunneling conductance (FITC)~\cite{S80}
has been proposed as a responsible mechanism in ZnO
nanowires~\cite{SQZE15}, in bundles of double-walled carbon
nanotubes~\cite{QELSMN15} and in nanoporous \tioo thin
films~\cite{KRSPBSB11}. The FITC model predicts non-linear
current-voltage $I-V$ curves due to intrinsic barriers between
grains in the sample~\cite{S80}. Other frequently used mechanisms
include variable-range hopping and thermally activated processes.
However, they cannot explain the non-linear behavior of $I-V$
curves~\cite{SQZE15} and the saturation of the resistivity at low
temperature~\cite{HCRNH03,BBFFMKPSTV05}.

In this work, procedures to isolate single \tioo nanotubes and to
establish barrier free contacts for electrical transport measurements
are presented.  The resistance of amorphous and polycrystalline
anatase samples were investigated in a broad temperature range. A FITC
mechanism contributes to the conductivity in all measured samples with
non-linear $I-V$ curves. Additionally, the electrical transport of a
highly insulating anatase nanotube was modified by means of defect
production at the surface using low-energy ion irradiation.

\section{Experimental}

The \tioo nanotubes have been separated from nanotubular layers
anodically grown on titanium foil (Advent Research Materials Ltd.,
99.6~\% purity) in an electrochemical cell with ethylene glycol
electrolyte containing 0.15~M ammonium fluoride and 1~M H$_2$O. The
titanium foil was anodized with a potential of 60~V applied for
5~hours. Finally, the nanotubular array was placed in ethanol and
dried with nitrogen gas.
\begin{figure}
\includegraphics[width=\columnwidth]{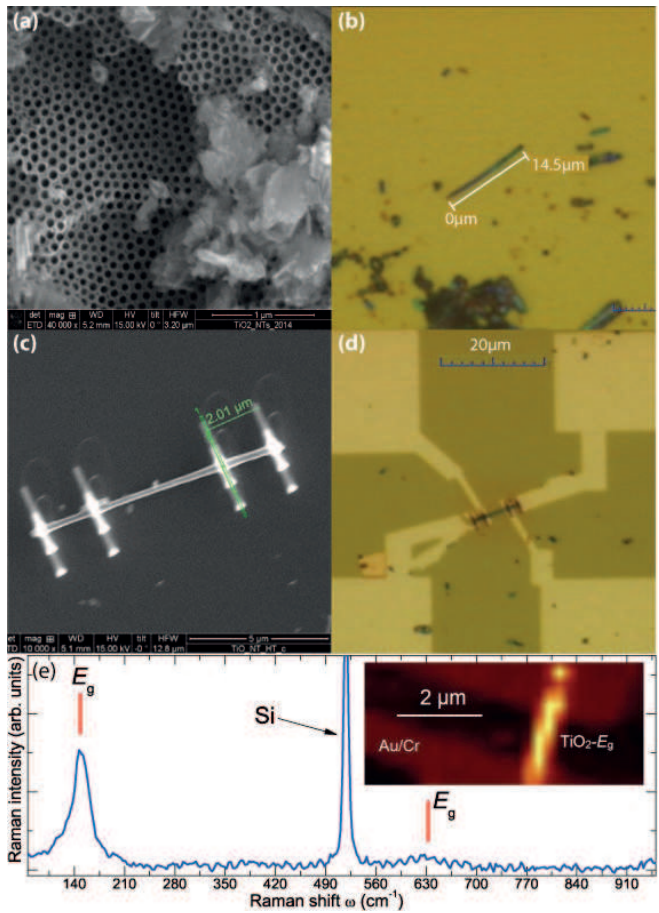}
\caption{\label{fig:fig1} In (a) a \tioo nanotubular array grown on Ti
  foil is shown. The nanotubes have an average diameter of
  approximately 120~nm. Figure (b) displays a single \tioo nanotube,
  in (c) the nanotube after WC$_x$ deposition and in (d) the contacted
  \tioo nanotube ready for measurement can be seen. In (e) a Raman
  image of the investigated sample is shown, the bright region
  corresponds to the Raman $E_g$ band at $148$~cm$^{-1}$.}
\end{figure}
The grown nanotubes (see \figurename~\ref{fig:fig1}(a)) are initially
in the amorphous phase and were annealed in air using a Rapid Thermal
Annealer (Jet-first Rapid Thermal Annealer, Jipilec, France) in order
to obtain polycrystalline anatase samples. The heating/cooling rate
was 15°$^\circ$C$/$min with annealing temperatures of 350$^\circ$C and
450$^\circ$C. \hl{After annealing, the nanotubes were polycrystalline
  and exhibited anatase morphology, for more detailed information
  please refer to Refs.}~\cite{ATFS10,LMS14,SHRYS16}. \hl{Annealing at
  higher temperatures results in mixed phases of anatase and rutile,
  eventually, the single nanotubes would
  collapse}~\cite{LMS14}. \hl{Therefore, only pure amorphous and
  anatase nanotubes were investigated.}

Bundles of nanotubes were scratched off the foil onto commercial,
p-boron doped silicon substrates ($5\times5$~mm) with a 150~nm SiN$_x$
coating, see \figurename~\ref{fig:fig1}(b).  By applying pressure and
a slow circular motion using a second substrate, some single nanotubes
\hl{break off the bundles} and were stuck to the substrates
because of electrostatic attraction.  Suitable tubes were selected
using an optical microscope. In order to fix the nanotubes on the
substrate and to prepare them for contacting, electron beam induced
deposition (EBID) of tungsten carbide was used, see
\figurename~\ref{fig:fig1}(c). The deposited WC$_x$ is nearly
insulating~\cite{SSRQFME07} and provides the necessary steps to
compensate the height difference between contacts and nanotube. The
substrates were covered with a positive working resist (ALL-RESIST,
PMMA 950~K, AP-R 671-05) and, by means of electron beam lithography
(EBL), the structures of the contacts were printed into the
resist. After developing, a bilayer film composed of Cr (5~nm) and Au
(35~nm) was sputtered. The PMMA was later removed by acetone. The
width of the contacts exceeds the WC$_x$ steps, in order to get a
potential barrier-free electrical connection. A prepared nanotube can
be seen in \figurename~\ref{fig:fig1}(d).  An overview of the
structural properties and dimensions of the nanotubes can be seen in
Table~\ref{tab:t1}.
\begin{table}
  \caption{\label{tab:t1} Overview of the \tioo nanotubes presented in this work. 
    The dimensions were measured using scanning electron microscopy. In order to 
    calculate the resistivity of sample NT4 after irradiation, a shell thickness of $d=5$~nm was assumed.}
\begin{tabularx}{\columnwidth}{@{}p{25pt}@{}p{42pt}@{}p{40pt}@{}X@{}p{80pt}@{}}\hline\hline
ID&Phase&Contacts&Length ($\upmu$m) &$\rho(T=300\mathrm{K})(\Upomega\mathrm{cm})$\\ \hline
NT1 &Anatase&4&$0.5\pm0.1$&0.026\\
NT2 &Anatase&4&$1.5\pm0.1$&0.044\\
NT3 &Amorph&2&$3.3\pm0.2$&5.31\\
NT4 &Anatase&2&$6.9\pm0.2$&34\\ \hline\hline
\end{tabularx}
\end{table}

For the transport measurements, each sample was contacted on a chip
carrier placed on the cold head of a standard closed cycle cooling
system inside a vacuum bell with a minimum temperature of
$T\approx 30$~K. The electrical resistance was measured using the
four-point probe configuration with a current source
(Keithley 6221) and a nano- voltmeter (Keithley 2182). The high
resistance measurements were performed with a constant applied voltage
using a DC source (Yokogawa 7651). The current was monitored with a
shunt resistance of $R_s=9.101~\mathrm{M}\Upomega$ in series with the
samples. For low temperature measurements down to $T=5$~K, a commercial
$^4$He cryostat (Oxford Instruments) was used.

The Ar ion irradiation was done in a self-made plasma chamber with a
parallel plate (copper) setup at room temperature. The chamber was
evacuated to a pressure of $P\approx0.1$~mbar with an Ar/H gas mixture
(Ar: 90~\% and H: 10~\%, Air Liquide) flowing through the chamber. The
chip carriers with samples were mounted $\approx12$~cm away from the
plasma center and a bias voltage of $U_{\rm bias}=50$~V was used to
accelerate the ions towards the sample, while connected to ground, and
the bias current was measured. \hl{The energy used is too low to
  produce any relevant sputtering, which could induce a composition
  variation.} Previously, the substrate was covered with PMMA and a
window was opened to shield the contacts using electron beam
lithography. The number of ions hitting the sample was estimated to be
$\approx2.2\times10^{14}$ Ar ions~\cite{SQESMLG16}.

Information about the sample structure was obtained using the confocal
Raman microscope alpha300R+ from WITec with an incident laser light of
wavelength $\lambda=532$~nm. The device has a lateral resolution of
$\approx300$~nm and a depth resolution of $\approx900$~nm. The energy
was kept at $\approx3$~mW to avoid damage caused by heating effects in
the sample.

\section{Results}

Using XRD, it was shown that the as-prepared nanotube bundles are
amorphous and they transform into anatase after
annealing~\cite{ATFS10,LMS14}. However, some single nanotubes could
remain in the amorphous state.  Using confocal Raman spectroscopy,
single isolated samples can be investigated. According to Ohsaka
\textit{et al.}~\cite{OIF78} for bulk anatase, Raman peaks can be
found at 639~cm$^{-1}$, 197~cm$^{-1}$ and 144~cm$^{-1}$, assigned as
$E_g$ modes. $B_{1g}$ modes are at 513~cm$^{-1}$ and 399~cm$^{-1}$, and
the band at 519~cm$^{-1}$ corresponds to the $A_{1g}$ mode.  The Raman
$E_g$ band at 144 cm$^{-1}$ is the most intense peak. Our results are
presented in \figurename~\ref{fig:fig1}(e), the peaks correspond to
the $E_g$ band, the obtained values are 148~cm$^{-1}$ and
633~cm$^{-1}$, respectively, which are different compared to the above
mentioned results for the bulk anatase.  This band shift is known as
blueshift and is caused due to phonon confinement in the
nanocrystals forming the nanotube.  This effect was already reported
for anatase nanocrystals~\cite{ZHZYC00}, where the shift of the $E_g$
peak as a function of the annealing temperature was investigated. A
blueshift to 148~cm$^{-1}$ was obtained for nanocrystals annealed at
$\approx350^\circ$~C, which is in agreement with the annealing
temperature used for the \tioo nanotube. From the blueshift, a
crystallite size of $\approx8$~nm could be obtained~\cite{ZHZYC00},
which is in agreement with XRD results~\cite{LMS14}. A $(x-y)$ Raman
scan is shown in the inset of \figurename~\ref{fig:fig1}(e), where the
bright shades correspond to the Raman $E_g$ band at 148~cm$^{-1}$.

The electrical properties of the nanotubes depend strongly upon the
phase and structural quality. A defect free \tioo anatase nanotubes is
electrical insulating. However, due to growth conditions, defects can
be introduced resulting in an electrical conductive material. For
example, such defects are oxygen vacancies (self doping) produced by a
reduction of \tio, e.g.~through electrochemical reactions, gas
annealing or exposure to vacuum~\cite{LMS14,LSFHVMSS14,PJLS12,GTMS06},
due to a separation of O$_2$ or H$_2$O from terminal oxide or
hydroxide groups and bridged oxide and Ti$^{3+}$ states~\cite{PJLS12}.
Many investigated nanowires and$/$or nanotubes exhibit non-linear
$I-V$ curves, which are usually neither discussed, nor explained in
the literature~\cite{CLL09}. Such non-ohmic behavior could be due to
intergrain conduction or barriers formed on the contacts used for
measurements.  For convenience, the measured samples in this work are
sorted in three categories: polycrystalline nanotubes with linear
(Section~\ref{sec:linear}) and non-linear
(Section~\ref{sec:nonlinear}) $I-V$ curves, an amorphous sample
(Section~\ref{sec:amorph}) and an insulating anatase nanotube which
was treated with low-energy Ar ions (Section~\ref{sec:ion}).

\subsection{Linear $I-V$ curves}
\label{sec:linear}

\begin{figure}[tb]
\includegraphics[width=\columnwidth]{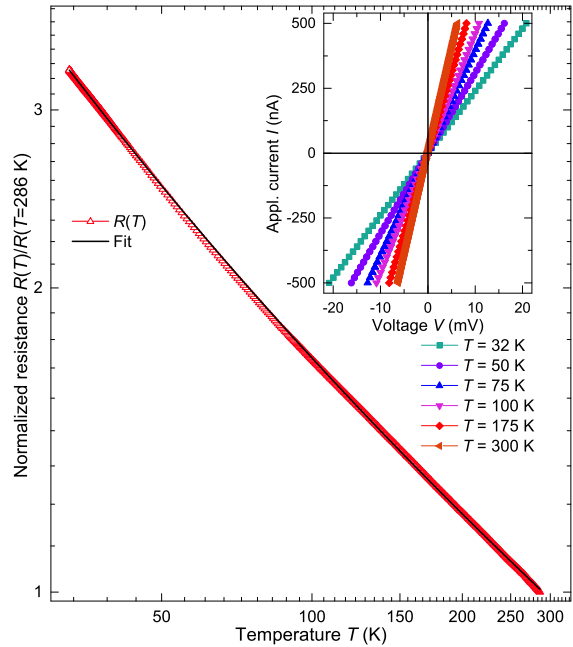}
\caption{\label{fig:fig2} The temperature dependent resistance of
  nanotube NT1 is shown. In the inset the linear $I-V$ curves can be
  seen. The continuous line was obtained from \mye{vrh1}.}
\end{figure}

The \tioo nanotube sample NT1 shows linear $I-V$ curves in the whole
investigated temperature range, see inset in
\figurename~\ref{fig:fig2}.  Similar results were already shown in
previous work~\cite{SQLEKAS13}. The measurements were carried out
using the four-point probes method. The temperature dependence of the
resistance can be seen in \figurename~\ref{fig:fig2}, which
can be fitted using the Mott variable range hopping (VRH)
model~\cite{SQLEKAS13}:
\begin{equation}\label{eq:vrh1}
R^{-1}_{\rm VRH}(T)=\left\{R_2\exp\left[\left(\frac{T_h}{T}\right)^{\nicefrac{1}{1+d}}\right]\right\}^{-1}+R^{-1}_0,
\end{equation}
where $R_2$ is an arbitrary prefactor, $R_0$ is a temperature
independent term, the dimensionality $d=3$ and $T_h$ is a
characteristic temperature. From the fit, \text{$T_h=(3450\pm29)$~K};
similar values have already been reported in the
literature~\cite{SQLEKAS13}. The density of states (DOS) at the Fermi
level $N(E_{\rm F})$ can then be calculated:
\begin{equation}\label{eq:vrh2}
  T_h=\frac{18}{k_{\rm B}\xi^3N(E_{\rm F})},
\end{equation}
where $k_{\rm B}$ is Boltzmann constant and the localization length is
assumed to be in the order of $\xi=1$~nm~\cite{CRWT97}. A DOS of
$N(E_{\rm F})\approx6.1\times10^{28}\mathrm{eV}^{-1}\mathrm{m}^{-3}$
was found, which agrees very well with the
literature~\cite{SQLEKAS13,YLKM07}. The resistivity, see
Table~\ref{tab:t1}, is very low compared to other investigated
samples. The low resistivity might be a consequence of doping due to a
large density of defects present in the sample.

\subsection{Non-linear $I-V$ curves}
\label{sec:nonlinear}

The current-voltage characteristics using Mott VRH and activated
transport processes correspond to linear $I-V$ curves (ohmic
regime).Therefore, they fail to explain non-linear $I-V$ curves as
well as $R(T)$ for such samples. The conduction of the polycrystalline
nanotube depends on the intragrain and intergrain conductivity. When
there is no doping, the grains are insulating with an energy gap of
$\approx3$~eV. At intermediate doping, the charge carriers move to the
crystal defects/boundaries between the grains, which are acting as
electronic traps, and thus a depletion layer is formed with a
potential barrier. In this case non-linear $I-V$ curves can be
observed also when measuring with four-point probes method. At high
doping levels, the material is saturated and the barrier vanishes
again.

\begin{figure}
\includegraphics[width=\columnwidth]{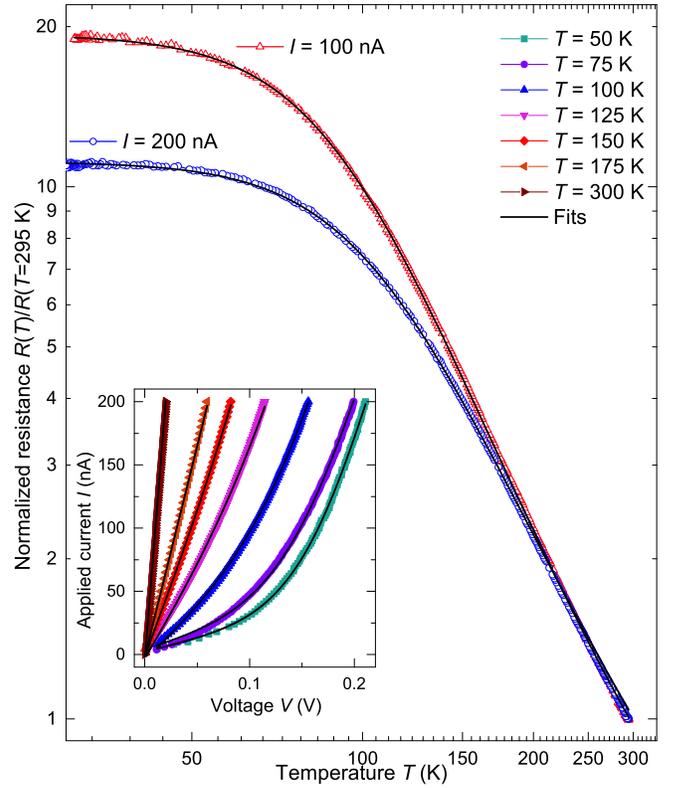}
\caption{\label{fig:fig3} The temperature dependent measurements at
  two applied currents of sample NT2 are presented.  The resistance at
  room temperature is $R_{\rm RT}~=~100~\mathrm{k}\Upomega$. The inset
  shows the $I-V$ curves at different temperatures. The continuous
  lines are the fits obtained following \mye{fitc1} for $R(T)$ and
  \mye{fitc2} for the $I-V$ curves.}
\end{figure}

The nanotube NT2 was measured using four contacts and shows non-linear
$I-V$ curves, see inset \figurename~\ref{fig:fig3}.
The $I-V$ measurements were performed from $T=50$~K to $T=300$~K,
where non-linear behavior can be observed at temperatures
$T\leq175$~K and below. This can also be seen in the temperature
dependent resistance measurements, which were done using $I=100$~nA
and $I=200$~nA, at $T\leq175$~K the curves split. The non-linearity
of the $I-V$ curves and the temperature dependence can be explained
using the fluctuation induced tunneling conductance (FITC) model,
which was already used to describe  similar materials such as
nanoporous \tioo thin films~\cite{KRSPBSB11}, ZnO
nanowires~\cite{SQZE15}, oxide
nanostructures~\cite{FLKUA07,LCL08,LL11}, double walled carbon
nanotube bundles~\cite{QELSMN15} or disordered
semiconductors~\cite{R75}.  According to the FITC model, at small
applied electric fields, the temperature dependent resistance across a
junction is given by~\cite{S80}
\begin{equation}
  \label{eq:fitc1}
  R^{-1}_{\rm FITC}=\left(R_\infty\exp\left[\frac{T_1}{T_0+T}\right]\right)^{-1}+R^{-1}_0,
\end{equation}
where $R_\infty$ is a free, temperature independent parameter, and the
characteristic temperatures are defined as
\begin{equation}
\begin{aligned}
T_1 &=\frac{8\epsilon_r\epsilon_0A\varphi^2_0}{e^2k_{\rm B}w},\\
T_0 &=\frac{16\epsilon_r\epsilon_0\hbar A\varphi^{3/2}_0}{\pi\sqrt{2m}k_{\rm B}e^2w^2},
\end{aligned}
\end{equation}
where $\epsilon_0$ is the vacuum permittivity, $\epsilon_r$ the
dielectric constant of the barrier, $e$ the elementary charge,
$k_{\rm B}$ is the Boltzmann constant, $\hbar$ the reduced Planck
constant, $m$ the electron mass, $A$ the area of the tunnel junction,
$\varphi_0$ the barrier height and $w$ is the barrier width. The
characteristic energy $T_1$ can be regarded as the energy required for
an electron to pass the barrier and $T_0$ is the temperature for which
well below thermal fluctuations become insignificant. As stated before,
the FITC model also provides the means to describe the non-linear
$I-V$ curves at different temperatures as follows~\cite{S80}:
\begin{equation}\label{eq:fitc2}
I_{\rm FITC}=I_s\exp\left[-a(T)\left(1-\frac{V}{V_c}\right)^2\right],|V|<V_c,
\end{equation}
where $I_s$ and $V_c$ are the saturation current and critical voltage,
respectively, and $a(T)$ is given as:
\begin{equation}\label{eq:aT}
  a(T)=\frac{T_1}{T_0+T}.
\end{equation}
As can be seen from these equations, the characteristic temperatures
can be obtained through fitting the \text{$I-V$} curves and the
temperature dependent resistance $R(T)$. However, in order to fit the
data, a temperature independent term $R_0$ in parallel to the FITC
conduction process, has to be added. The parallel contribution is
due to disorder and impurities present in the \tioo nanotubes.

In order to fit the data and to reduce the amount of free parameters,
all curves were fitted simultaneously ($I-V$ curves and $R(T)$) and
the corresponding parameters were taken as shared parameters for all
curves. This means, that $V_c$ and $I_s$ (which depend only weakly on
the temperature) are shared among the data of the $I-V$ curves, and
that $T_1$ and $T_0$ are shared among the $I-V$ curves and $R(T)$
results. The data and the fits can be seen in
\figurename~\ref{fig:fig3}, the FITC model describes very well both
the $I-V$ curves and $R(T)$ results.  From the fit results, a
saturation current of $I_s\approx2.8\times10^{-7}$~A and a critical
voltage of $V_c\approx0.29$~V are obtained. The characteristic
temperatures are $T_1\approx853$~K and $T_0\approx59$~K. Similar
values have already been reported in the
literature~\cite{LCL08,LL11,XS09,SQZE15}. Although, the samples were
measured using the four-point probes method, non-linear $I-V$ curves
were measured as consequence of barriers formed at the intergrain
boundaries. This effect could be avoided by employing long term
annealing at intermediate temperatures, as high-temperature annealing
would result in a collapse of the \tioo
nanotubes~\cite{LMS14,AGADLTMS08,MASLM15,SHRYS16}.

\subsection{Amorphous nanotube}
\label{sec:amorph}

\begin{figure}
\includegraphics[width=\columnwidth]{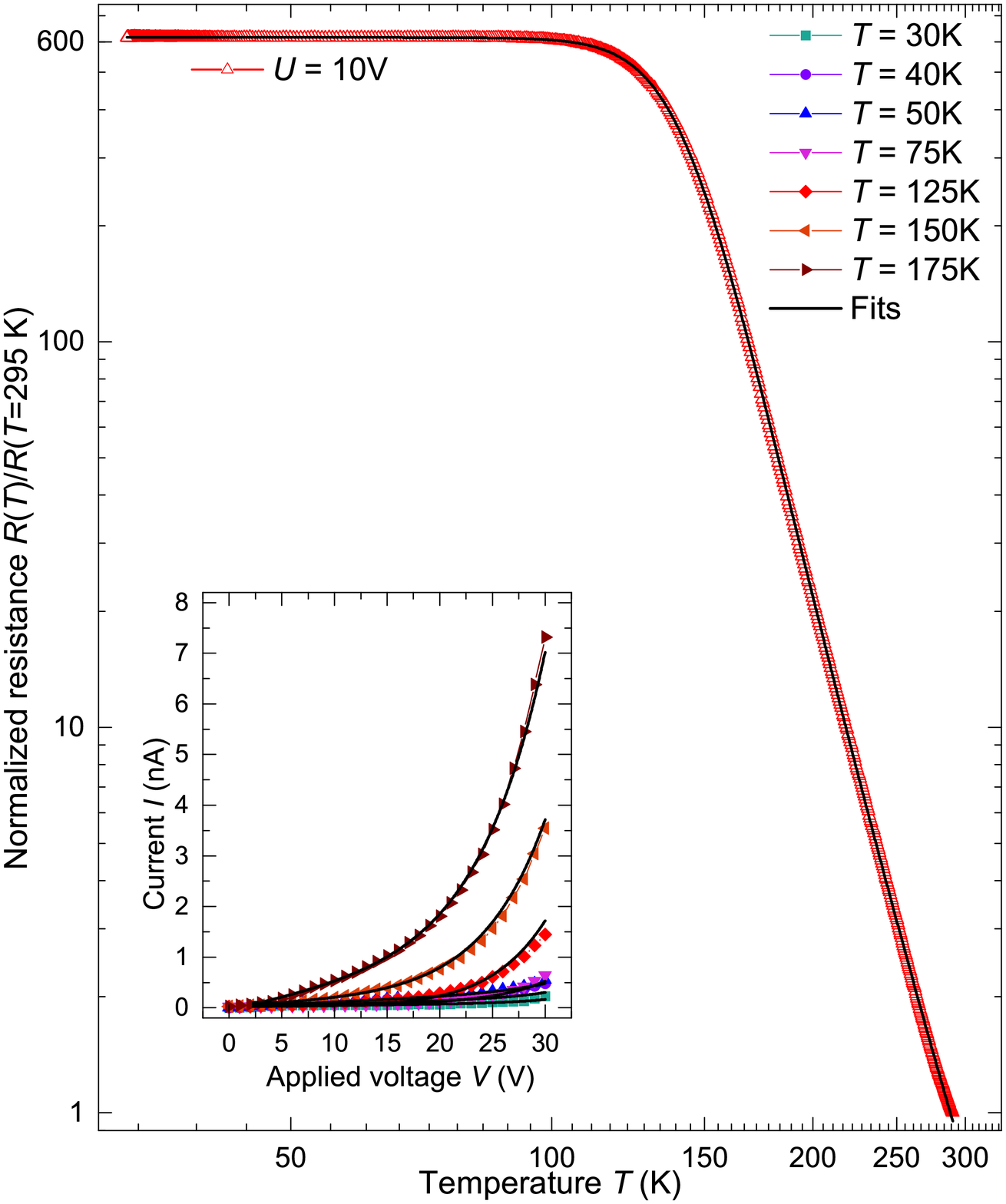}
\caption{\label{fig:fig4} The temperature dependent measurements at
  constant applied voltage of sample NT3 are presented.  The
  resistance at room temperature is
  $R_{\rm RT}~=~29~\mathrm{M}\Upomega$. The inset shows the $I-V$
  curves at different temperatures.  The continuous lines are the fits
  obtained following \mye{vrh1} and \mye{fitc1} for $R(T)$, and
  \mye{fitc2} for the $I-V$ curves.}
\end{figure}

The temperature dependent resistance  of an amorphous \tioo
nanotube NT3 is shown in \figurename~\ref{fig:fig4}.
The room temperature resistance is
$R(295~\mathrm{K})=29~\mathrm{M}\Upomega$, and thus much higher than
the previously shown samples. Therefore, the resistance was measured
with a constant applied voltage of $V=10$~V and the current was
monitored with a shunt resistance. This implies a two point-probes
technique to be used. However, a large influence of the contacts is
not expected, due to the very high resistance of the \tioo nanotube
itself. This assumption is supported by the result of a four-point
probes measurement at room temperature, which yields the same
resistance as for the two-point measurement. Therefore, the influence
of the contacts will be neglected. The data as well as the fits are
shown in \figurename~\ref{fig:fig4}, in the inset the $I-V$ curves can
be seen. As before, all measurements were fitted simultaneously. In
order to fit the $R(T)$ data, not only the FITC model has to be
assumed but also a VRH hopping contribution in parallel was needed,
see \mye{vrh1}. The shared parameters were:
$I_s=5.5\pm0.1\times10^{-5}$~A, $V_c=144\pm3$~V, $T_1=6545\pm50$~ and
$T_0=272\pm7$~K. With the characteristic temperature
$T_h\approx70000$~K and using \mye{vrh2}, the DOS at $E_{\rm F}$ is
$N(E_{\rm F})\approx3\times10^{27}$~eV$^{-1}$m$^{-3}$. This value is
one order of magnitude smaller than what was obtained for \tioo
nanotube with linear $I-V$ curves. This together with the large values
of $T_1$ and $T_0$, i.e.~large barrier height, explain the high
resistance of this sample. At low temperature, the constant $R_0$ term
in parallel which is due to impurities$/$defects, dominates the transport.

\subsection{Ar ion irradiated nanotube}
\label{sec:ion}

\begin{figure}
\includegraphics[width=\columnwidth]{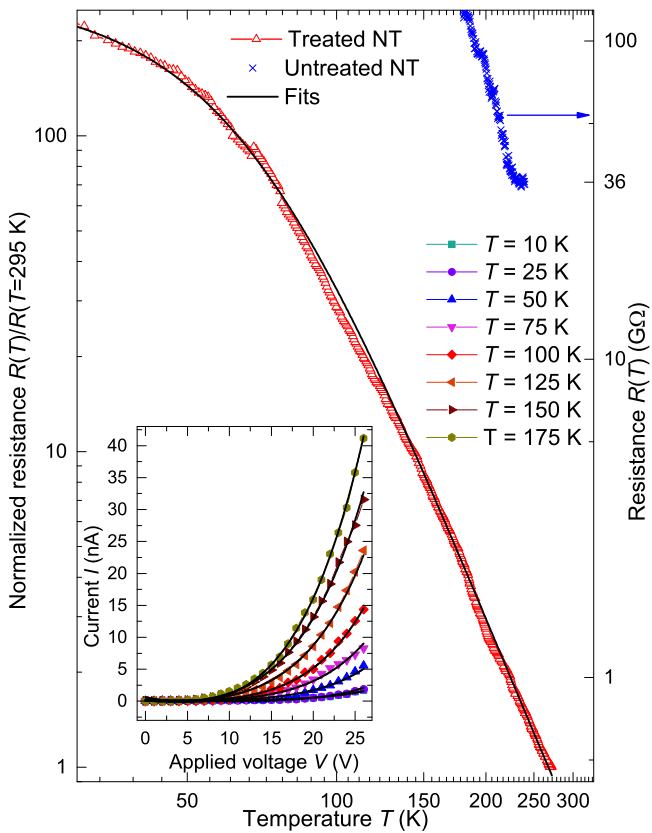}
\caption{\label{fig:fig5} The results of the temperature dependent
  measurements at constant applied voltage of sample NT4 after Ar$/$H
  ion irradiation are presented.  The resistance at RT is
  $R_{\rm RT}~=~500~\mathrm{M}\Upomega$. The inset shows the $I-V$
  curves at different temperatures. The continuous lines are the fits
  obtained following \mye{fitc1} for $R(T)$ and \mye{fitc2} for the
  $I-V$ curves. The resistance before ion irradiation was
  $R_{i}~=~36~\mathrm{G}\Upomega$ at $T=240$~K and is indicated
  with blue crosses.}
\end{figure}

In order to investigate the influence of defects on the transport
properties of polycrystalline anatase \tioo nanotubes, an almost
insulating sample was chosen (see results in
\figurename~\ref{fig:fig5}), indicating a high quality of the
crystalline structure. The sample NT4 has been irradiated using an
Ar/H plasma. The results after irradiation of the temperature
dependent resistance measurements and $I-V$ curves can be seen in
\figurename~\ref{fig:fig5} and its inset.
The used energy of the plasma ions of $50$~V and, according to SRIM
simulations, the resulting penetration depth is~$\approx5$~nm,
implying that the nanotube is modified only at the
surface~\cite{SQESMLG16,SQZE15}. The sample consists then of an
insulating polycrystalline nanotube surrounded by a conducting \tioo
shell. The FITC model is suitable to describe such systems, which is
in agreement with the non-linear $I-V$ curves. Although the sample was
irradiated homogeneously, the conduction can be well described by the
FITC model. This indicates that the defects produced in the grains
induce finite conductivity, yet the regions between the grains,
i.e.~the grain boundaries, remain insulating and are less affected by
defects.  The sample was measured using the two-point probes method
and shunt resistance. Due to the high intrinsic resistance, the
contact contribution is neglected. The $I-V$ curves and $R(T)$ were
fitted as before, using the characteristic temperatures $T_1$ and
$T_0$ and the saturation current $I_s$ and critical voltage $V_c$ as
shared parameters. The model to fit the data was the same as for the
other polycrystalline sample with non-linear $I-V$ curves, i.e.~the
FITC model in parallel with a temperature independent residual
resistance. The obtained parameters are
$I_s=7.9\pm0.6\times10^{-6}$~A, $V_c=116\pm10$~V, $T_1=3106\pm80$~ and
$T_0=206\pm4$~K. This indicates a larger barrier height compared to
the other polycrystalline samples. The estimated resistivity, see
Table~\ref{tab:t1}, is larger than of sample NT2, indicating a
relative low defect production probability with the used energy.

\section{Conclusion}

Several anodically grown amorphous and polycrystalline \tioo nanotubes
were isolated and prepared for the measurement of their electrical
transport properties. Raman spectroscopy reveals that the investigated
anatase samples are homogeneous and polycrystalline with a grain size
of a few nanometers.  For nanotubes with linear $I-V$ characteristics
a VRH transport mechanism explains the measured behavior.  In order to
describe the $R(T)$ and non-linear $I-V$ curves, the FITC as well as
the VRH model are used. Using four contacts, non-linear $I-V$ curves
were measured, which can be explained considering a barrier
formed at the interfaces between the grains.  The fluctuation induced
tunneling conductance describes the resistance results as well as the
non-linear $I-V$ curves for the polycrystalline \tioo nanotubes. A
combination of the FITC model and VRH was used for the analysis of the
resistance of an amorphous nanotube. The contacts on the crystalline
samples are ohmic, i.e.~there is no barrier, which is
important for future studies and applications.  An insulating sample
was irradiated with low-energy Ar/H plasma, and a large change in the
resistivity was produced. This provides the possibility to modify the
electrical transport properties of individual \tioo nanotubes through
controlled irradiation with ions. In this work, the preparation of
single \tioo nanotubes with ohmic contacts for electrical transport
measurements was demonstrated, which opens new possibilities for
future applications.

\begin{acknowledgments}
  This work was supported by DFG through the Collaborative Research
  Center SFB 762 ``Functionality of Oxide Interfaces''.
\end{acknowledgments}

\section*{Bibliography}
\bibliography{bibliography}

\end{document}